\documentclass[authorversion,screen]{acmart} 
\usepackage{graphicx} 
\usepackage{subcaption}
\usepackage{caption}

\acmConference[SCALLY-25 at CHI '25]{Workshop on Scaling Distributed Collaboration in Mixed Reality (SCALLY-25)}{May 2025}{Yokohama, Japan}
\acmBooktitle{Proceedings of SCALLY-25 at CHI '25, May 2025, Yokohama, Japan}
\acmYear{2025}
\acmDOI{10.1145/0000000.0000000}
\acmISBN{978-1-4503-XXXX-X/20/XX}



\setcopyright{rightsretained}

\newcommand{\figArtifact}{
\begin{figure}
  \centering
  \includegraphics[width=0.9\textwidth]{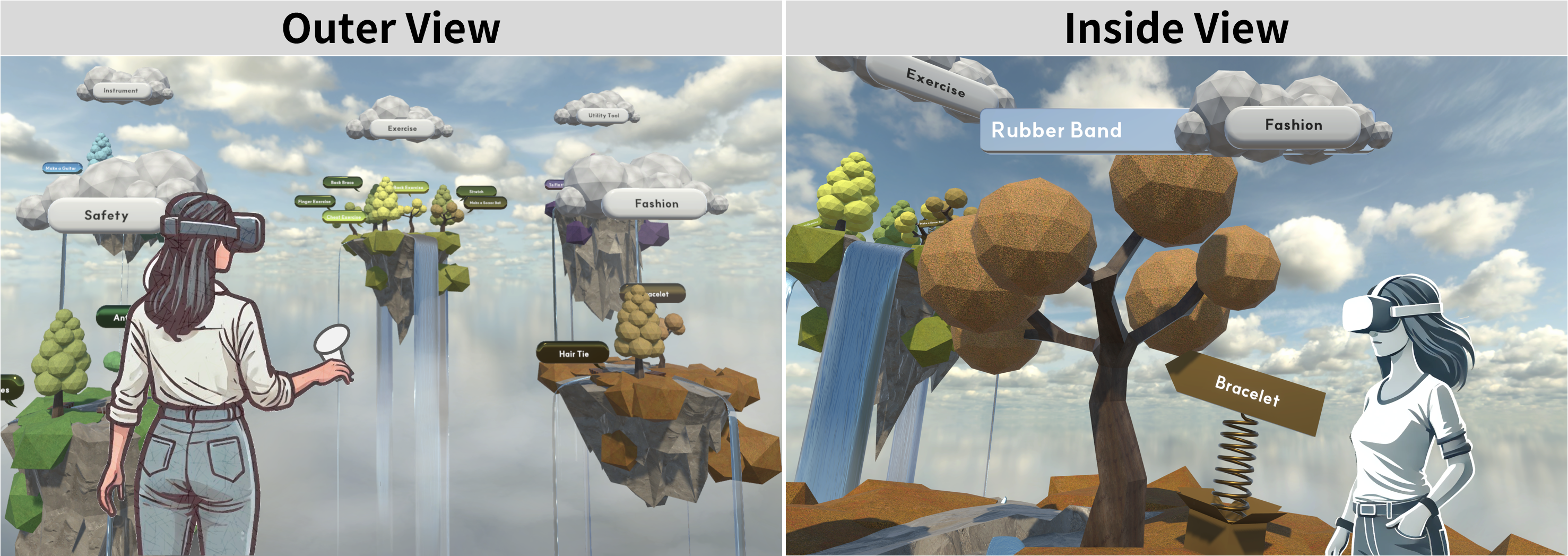}
  \caption{Overview of \textit{Idea Islands}: a system that virtualizes individual ideation by mapping ideas into interactive metaphors, supporting dynamic in-metaphor exploration.}
  \label{fig:artifact}
\end{figure}
}

\newcommand{\figComponent}{
\begin{minipage}{0.57\textwidth}
    \centering
    \includegraphics[width=\textwidth]{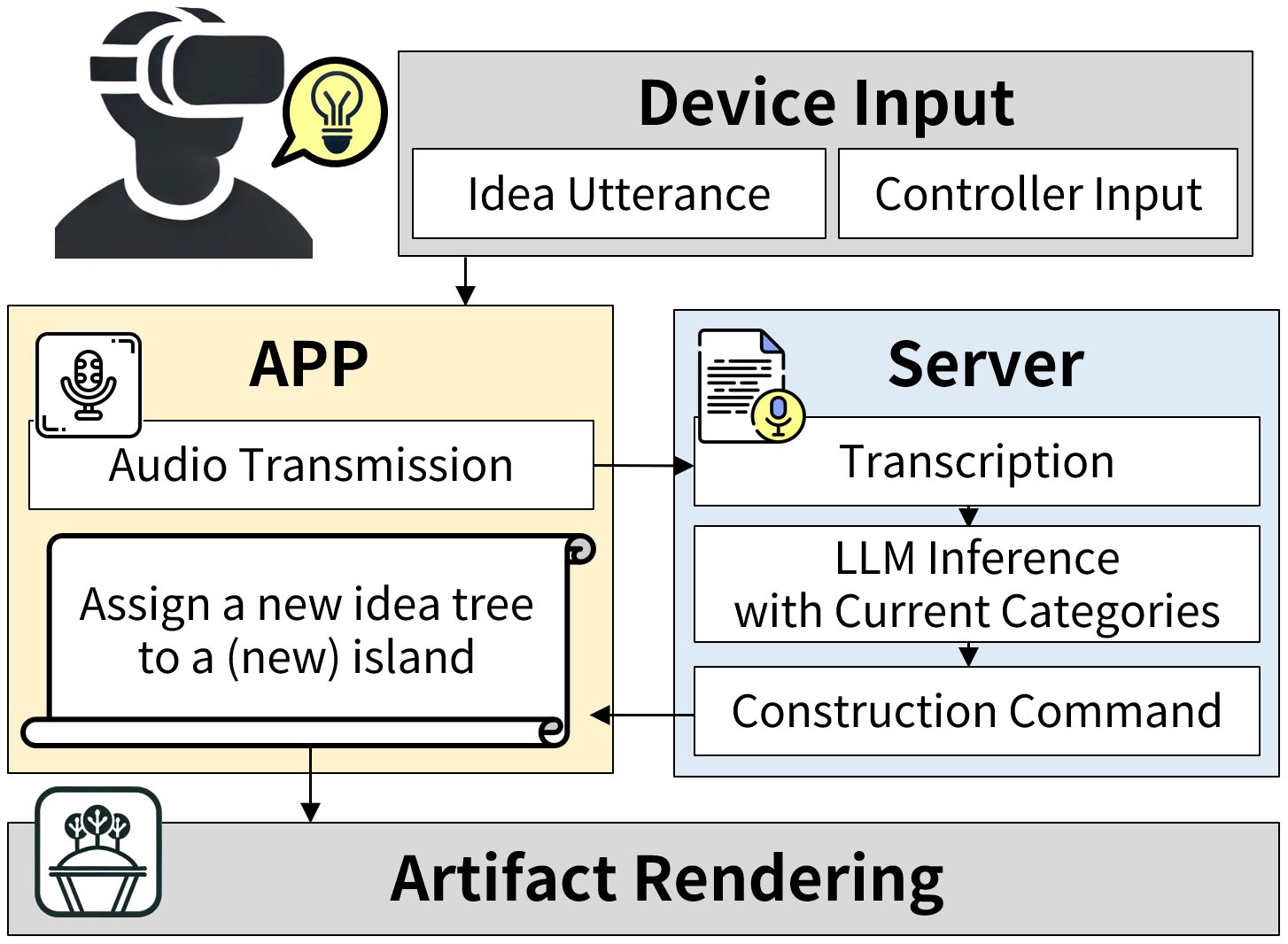}
    \captionof{figure}{The transformation of ideation into an interactive virtual task space using \textit{Idea Islands}.}
    \label{fig:pipeline}
\end{minipage}\hfill
\begin{minipage}{0.38\textwidth}
    \centering
    \includegraphics[width=\textwidth]{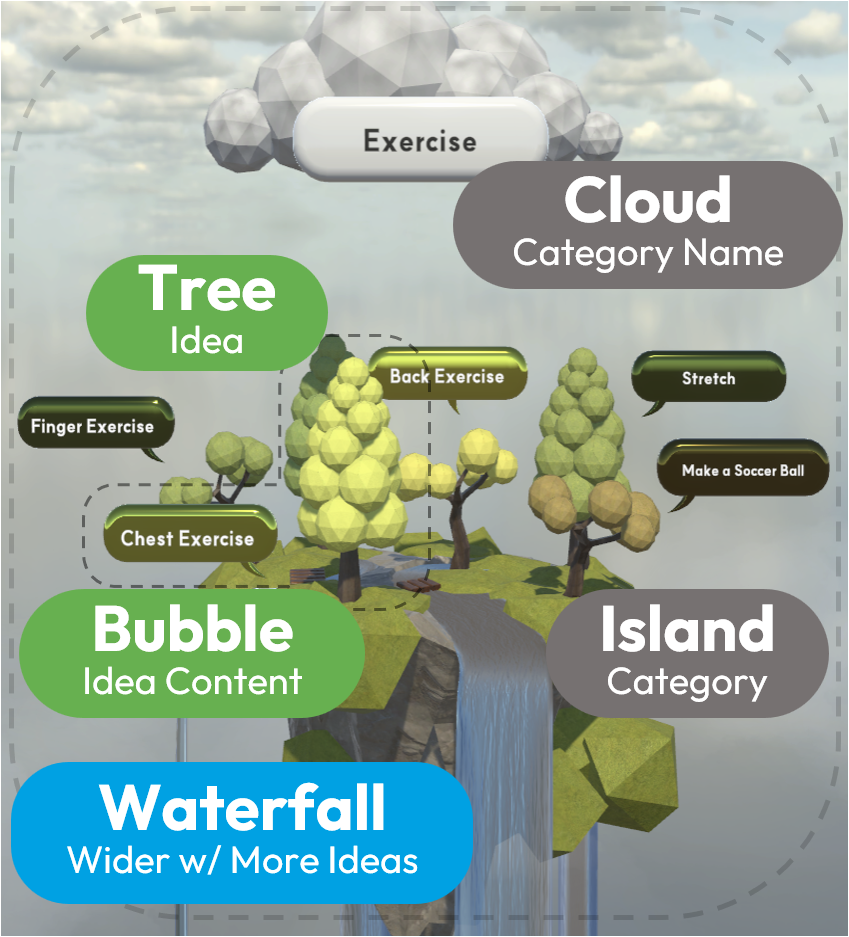}
    \captionof{figure}{Key elements of an \textit{Idea Island}.}
    \label{fig:elements}
\end{minipage}
}

\title{Virtualizing a Collaboration Task as an Interactable Environment and Installing it on Real World}

\author{Euijun Jung}
\affiliation{%
  \institution{Human-centered Computer Systems Lab, Seoul National University}
  \country{Republic of Korea}
}
\email{euijun.jung@hcs.snu.ac.kr}
\author{Youngki Lee}
\affiliation{%
  \institution{Human-centered Computer Systems Lab, Seoul National University}
  \country{Republic of Korea}
}
\email{youngkilee@snu.ac.kr}

\received{14 March 2025}
\received[accepted]{24 March 2025}

\begin{document}
\begin{abstract}
This paper proposes a novel approach to scaling distributed collaboration in mixed reality by virtualizing collaborative tasks as independent, installable environments.
By mapping group activities into dedicated virtual spaces that adapt to each user's real-world context, the proposed method supports consistent MR interactions, dynamic group engagement, and seamless task transitions.
Preliminary studies in individual ideation demonstrate enhanced immersion and productivity, paving the way for future multi-user collaborative systems.
\end{abstract}

\maketitle

\section{View on Topics}

The workshop outlines six key topics:
(1) creating congruent MR interaction spaces across diverse physical environments; (2) accurately portraying embodied interactions across distributed teams; (3) supporting dynamic group interactions within large, distributed MR settings; (4) enabling effective asynchronous collaboration; (5) supporting shared interactions with tangible objects across distributed locations; and (6) supporting seamless transitions between collaborative tasks.

I propose that group activities can be conceptualized as a single, interactive virtual space that encapsulates the task.
By defining a group activity as its own virtual space and then ``installing" it to align with each user’s real environment, distributed collaboration can be effectively scaled.

The following sections outline this approach, present preliminary studies, and discuss potential developments aligned with the workshop’s theme of scaling distributed collaboration.

\section{Virtualization of Collaborative Task as Environment}

Virtualizing a collaborative task as an interactive environment addresses the six aspects outlined in the workshop.

\paragraph{Creating congruent MR interaction spaces across diverse physical environments}  
By defining a group activity as an independent virtual space and \textit{installing} it within each user’s actual environment, this approach provides a consistent MR interaction space across different physical settings.
In straightforward cases, where users access the virtual space from an empty, large enough area, the virtual content can be directly scaled to match the available space.
For users in environments with persistent physical objects (e.g., desks, chairs) that must remain in interaction, a virtual proxy of these objects can be mapped onto the existing ones \cite{seo2024gradualreality}.
In environments where ongoing interaction is required for safety (e.g., while walking or using public transportation) and the task is less directly relevant, task-related information can be concisely represented—using, for instance, a semantic task dependency graph—to balance focus between the physical and virtual realms.
Moreover, the degree of virtual representation can be dynamically adjusted according to the user’s cognitive load and level of task immersion \cite{lindlbauer2019context}.

\paragraph{Accurately portraying embodied interactions across distributed teams}  
Within a dedicated virtual space, the essential interactions, gestures, and movements of group activities can be captured and faithfully reproduced.
The level of detail in these embodied interactions can be varied based on the user’s requested level of detail when accessing the virtual space.
For instance, in a fully immersive task space, high-fidelity avatars or realistic human representations may be employed \cite{guan2023metastream, lee2023farfetchfusion, cheng2024magicstream}, whereas for lighter interactions, displaying only facial features or even omitting avatars might suffice.
Note that this aspect is orthogonal to the core proposal.

\paragraph{Supporting dynamic group interactions within large, distributed MR settings}  
By designing the virtual task space in a modular way, it can adapt flexibly and in real time to large-scale distributed MR environments.
Once the group collaboration is defined as a dedicated task space, sub-tasks can be provided as different areas within that space, according to its operational hierarchy supporting various subgroup interactions (as separated \textit{island} spaces in Section \ref{sec:representation}).
This bases the dynamic creation of spaces to meet emerging needs, providing an immersive environment for all relevant participants.

\paragraph{Enabling effective asynchronous collaboration}  
Defining the virtual space as an independent workspace allows users to access and contribute from their own environments regardless of time constraints, supporting effective asynchronous collaboration.
Individual contributions can be reified as objects within the space, allowing progress to be visually confirmed over time (as \textit{tree} generation in Section \ref{sec:representation}).
This ensures that even participants who are not present concurrently can track the overall evolution of the task and plan their contributions accordingly.
When changes occur in a participant’s absence, mechanisms such as commit messages, action replays, or LLM-generated summaries can support effective catch-up, further enhancing asynchronous collaboration.

\paragraph{Supporting shared interactions with tangible objects across distributed locations}  
For shared objects, a virtual counterpart and its functions can be defined.
It is feasible to map the traits of this virtual object in the virtual task space onto real-world objects \cite{han2023blendmr, hettiarachchi2016annexing, dogan2024augmented}.
This mapping can enhance work efficiency and deepen collaborative immersion.
In turn, contextualized pop-ups, visual effects, notifications, and interactions can be supported in the physical environment, reflecting the unique context of the corresponding physical object.

\paragraph{Supporting seamless transitions between collaborative tasks}  
By defining task spaces for each collaborative task and mapping them onto the physical environment, the task space functions as a ``program" while the physical space serves as ``physical memory."
Transitions between these task spaces can be managed similarly to context switching. 
If the fidelity of the task representation or the utility of the physical space are sufficiently low, multiple task spaces can be simultaneously mapped in an ambient manner \cite{ishii1997tangible} within the physical environment and triggered to expand as needed.

\section{Representation of Virtual Task Space}
\label{sec:representation}
\figArtifact
\figComponent
Implementing a virtual task offers many benefits for group collaboration.
In this section, I present preliminary research on how such a space can be realized.

Ideation is one of the most fundamental and critical activities in group collaboration \cite{paulus2003group}.
Due to its complexity stemming from multiple participants \cite{nijstad2006group}, I first explored how to configure the task space for individual ideation.
For ideation to yield creative outcomes, it is important to organize existing ideas clearly \cite{finke1996creative} and engage in reflection-in-action \cite{schon2017reflective} to generate further insights.
Visualizing the ideation process is known to enhance such situational awareness \cite{siangliulue2016ideahound, chandrasegaran2019talktraces} and engagement \cite{tausch2016comparison, tausch2014groupgarden}.
Additionally, incorporating walking activities during ideation boosts creative thinking \cite{oppezzo2014give}.

\textit{Idea Islands} is an artifact designed to promote ideation by visualizing the ideation context in real time while supporting walking activities within that space (Figure \ref{fig:artifact}).
To transform ideation into a ``task space," Idea Islands forms \textit{explorable metaphors}, adopting the metaphor of islands and trees.
When a user remarks an idea, the system infers its category and automatically creates an ``island" corresponding to that category (Figure \ref{fig:pipeline}), within which a ``tree" representing the idea is generated (Figure \ref{fig:elements}).
This metaphorical representation conveys the overall context of ideation to the user, who can navigate the categorized islands and interact with objects such as the trees within them.

Comparative experiments conducted under various conditions—such as the viewpoint of the metaphor (outside view versus inside view), the method of generating the metaphor (separating islands by themes or mapping ideas to a single island), and the mode of construction (automatic versus manual)—confirmed that these factors significantly influence the ideation process.
For instance, the separation of idea categories increased the number of ideas by 17.2\% and idea diversity by 26\% compared to a baseline condition, while single island mapping increased the number of ideas by 25.8\% but not the diversity.

The influence of the metaphor on participants can be analyzed into three main factors. 
First, the visual elements and their playful effects enhanced users’ immersion and attention, increasing their engagement.
Second, the way the metaphor (whether separating islands) is constructed shifts the emphasis on different elements, affecting the ideation strategies.
Third, the mechanisms by which users interact with the metaphor (manually or being inside metaphors) change the points of focus and interest of experience.

Based on these experimental results, follow-up studies are currently underway to enhance walking activities within the space and to further strengthen the characteristics of spatialization.

\section{Installation of Virtual Task Space on the Real World}

The artifact supports individual ideation, and it can be extended to facilitate collaboration by incorporating input from multiple users and placing outputs and user avatars in the same task space.
Once expanded, the system can evolve in various ways.
One promising direction is to support tasks of more complex structures than simple ideation.
The artifact would dynamically reflect each participant's progress regarding the task in real time, and it could be installed within the physical spaces where each participant is located.
Taking the above artifact as a reference, imagine mapping a specific category—represented as an ``island" that you wish to further develop—onto your current physical space (for example, your room).
In this mapped area, objects such as ``trees" or ``waterfall" would become interactive elements within your environment.
Interactions with such objects are expected to support a more sustained and immersive engagement with the ongoing task.

To enable this installation, several key aspects need to be explored:
\begin{enumerate}
    \item Definition of the portable structure of task: Develop a method transforming a target task into an adaptively installable task space structure reflecting the task's objectives.
    \item Abstraction of the real-world environment: Create an abstraction that considers the semantics and spatial arrangements of objects in the physical environment.
    \item Effective matching strategy: Establish a method for aligning the virtual task space with the physical space.
    \item Interaction mechanism deployment: Determine natural yet versatile interaction mechanisms for the target physical context, supporting user immersion and productivity.
\end{enumerate}

Based on these investigations, it is anticipated that scalable collaboration can be achieved through the independent configuration of task spaces and their tailored installation in individual physical environments.

\bibliographystyle{ACM-Reference-Format}

\end{document}